**LETTER TO THE EDITOR**

**Open Access**

# Missed opportunities in large scale comparison of QSAR and conformal prediction methods and their applications in drug discovery

Damjan Krstajic[*]

## Abstract

Recently Bosc et al. (J Cheminform 11(1): 4, 2019), published an article describing a case study that directly compares conformal predictions with traditional QSAR methods for large-scale predictions of target-ligand binding. We consider this study to be very important. Unfortunately, we have found several issues in the authors' approach as well as in the presentation of their findings.

## Introduction

Recently Bosc et al. [1] published an article in the *Journal of Cheminformatics*, describing a case study that directly compares conformal predictions with traditional QSAR methods for large-scale predictions of target-ligand binding. We consider this study to be very important. Unfortunately, we have found several issues in the authors' approach as well as in the presentation of their findings. We shall start with more general issues and then go on to more specific ones.

## Generalisation of QSAR vs MCP

The authors' aim was to compare QSAR and Mondrian conformal prediction (MCP) as objectively as possible, making full use of all the data, subject to the constraints inherent in each method. QSAR classification models were built using the Random Forest (RF) with the number of trees and the maximum depth of the tree set to 300 and 20 respectively. All other RF parameters were set to their default values. The authors inform us that their internal tuning experiments using grid search demonstrated that these values generally enabled them to obtain the most accurate models (data not shown). Similarly, the same RF algorithm and associated parameters were used in the MCP framework. This means that whenever the authors refer to QSAR and MCP models they are de facto discussing RF with predefined fixed parameters as a QSAR model and its implementation in the MCP framework.

After a brief description of how RF was used in the paper, the authors then continue to present and discuss the results throughout the paper as comparisons between QSAR and MCP models, without ever mentioning again that they are really analysing the results of an RF model with fixed parameters in QSAR and MCP settings.

We understand that in order to analyse datasets on a such large scale the process needs to be simplified. Therefore it might not be feasible to apply other model building techniques as well as model selection processes [2] for each dataset separately. However, in our opinion that cannot be the reason, nor do the authors provide any explanation, for generalising their findings in the paper (most importantly in the title, abstract, graphs, tables and conclusion) as QSAR vs MCP models. Consequently, we do not consider such extrapolations as beneficial to research regarding the comparison of QSAR and MCP models.

*Correspondence: damjan.krstajic@rcc.org.rs
Research Centre for Cheminformatics, Jasenova 7, 11030 Belgrade, Serbia

## Variability

When performing a comparison between two methods, in our view, it is very important to address the issue of the variability of generated results. The paper does not tell us whether the authors have addressed the following causes of variability, and if so how. There is also some information missing in their presentation of variability in the paper (details in the subsection Presentation of variability).

### Random forest variability

If we repeat the process of building a logistic regression model (linear model) and applying it on a dataset, we shall obtain identical predicted probabilities and predicted labels. However, this is not guaranteed with RF models. As its name suggests there is randomness inherent in RF models. Therefore, unless we fix a random seed number, if we re-build and re-apply the RF model we will obtain somewhat different predicted probabilities. The predicted labels may or may not be identical. It depends on samples with predicted probabilities close to 0.5. The randomness may affect their predicted probabilities to flip-flop around 0.5 and thus change their predicted labels. We could not find that the issue of variability caused by RF in QSAR and MCP settings is either considered or addressed in the paper.

### MCP variability

In the Model validation section of the paper (page 5), the authors explain how for each target they split the dataset into a training (80%) and a test set (20%). For MCP they further randomly divided the training dataset into a proper training set (70%) and a calibration set (30%). Therefore for QSAR we would have the training set (80%) and the test set (20%), while for MCP we would have the proper training set (56%), the calibration set (24%) and the same test set (20%). Our understanding from the paper is that the authors have repeated 100 times the procedure of splitting the dataset into training (80%) and test (20%), and we agree with that. However, it is not mentioned in the paper, and therefore we are left to presume, that the splitting of each training set into the proper training and the calibration set was done only once. That is understandable because that is how MCP works. Nevertheless, if our aim is to compare QSAR with MCP models then we would have liked to see the variability of the MCP results caused by different splitting of the training set into the proper training set and calibration set, if not analysed, then at least mentioned in the paper.

### Presentation of variability

Throughout the paper the values of various statistics are presented with ± another value next to it. We presume that the other value is the standard error but we are not informed about it. Furthermore, we are not informed regarding the sample size from which the standard error was calculated. Our understanding is that the average values presented in the paper are not all calculated from samples with equal sample size. In Fig. 3 and Fig. 4 the variabilities of results are visually presented without the information they represent. Similarly in Table 2 and Table 4 next to each statistic there is ± of a value in the brackets without an explanation as to what they represent. It would have been nice if the authors had explained the variabilities presented in the paper and especially in tables and graphs, which we believe is common practice.

## Model validation

There is an important section of the paper which we have trouble understanding. In the first paragraph of the Model validation section (page 5), after a very understandable explanation of how the dataset was split into a training and a test set, and how then for MCP the training dataset was further split into a proper training and calibration set, we have the following sentence:

> "The splitting procedure was repeated 100 times using the different random splits and the result for each compound was obtained by calculating the median probabilities for QSAR or p values for MCP, over the 100 predictions."

The second part of the above sentence is confusing. We would like to remind that the authors at the start of the Model building section of the paper (page 4) state "*We chose to build simple active/inactive classification models.*". Therefore, we have so far been dealing with binary classifications. Our understanding is that the splitting into training (80%) and test set (20%) was done 100 times, which means that we would have 100 times 20% of compounds with binary predictions. So, where do "*the median probabilities for QSAR*" come from? Also "*the result for each compound was obtained …, over the 100 predictions*" as if each compound had 100 predictions!? The entire second part of the sentence in question does not make any sense to us. We would like to point out that this is in our opinion the crucial section for understanding the process of how the authors generated results for comparing QSAR and MCP models. We apologise in advance if this section is clear to others, and not to us.

## Explanation of _excl and _incl statistics

In order for us to explain other issues that follow, it is vital to understand the six new metrics introduced by the authors. In CP, binary classifiers may return the following four predictions: {positive}, {negative}, {positive and negative} and {null}. Authors referred to the last two predictions as 'both' and 'empty'. They introduced three metrics (sensitivity_incl, specificity_incl and CCR_incl) when compounds assigned to the 'both' class are considered correctly classified, and three further metrics (sensitivity_excl, specificity_excl and CCR_excl) where compounds in the 'both' class are ignored.

In our opinion it is worth examining the six new metrics in detail and we provide below additional detailed explanations for them in our own words. In the Table 1 we show the distribution of MCP predictions depending on their true value and the names we have ascribed to them.

$$TOTAL\_P = TP + FN + BP + EP - \text{total number of positives}$$

$$TOTAL\_N = TN + FP + BN + EN - \text{total number of negatives}$$

It was clarified to us that equations for 6 new metrics are then as follows:

$$Specificity\_excl = TN/TOTAL\_N,$$

$$Sensitivity\_excl = TP/TOTAL\_P$$

$$CCR\_excl = (specificity\_excl + sensitivity\_excl)/2$$

$$Specificity\_incl = (TN + BN)/TOTAL\_N$$

$$Sensitivity\_incl = (TP + BP)/TOTAL\_P$$

$$CCR\_incl = (specificity\_incl + sensitivity\_incl)/2$$

**Table 1 The distribution of MCP predictions depending on the actual value**

| Actual\MCP prediction | Positive | Negative | Both | Empty |
|---|---|---|---|---|
| Positive | TP | FN | BP | EP |
| Negative | FP | TN | BN | EN |

*TP* number of true positives, *FP* number of false positives, *TN* number of true negatives, *FN* number of false negatives, *BP* number of positives predicted as 'both', *EP* number of positives predicted as 'empty', *BN* number of negatives predicted as 'both', *EN* number of negatives predicted as 'empty'

### 'Both' predictions

We question the authors' analysis of 'both' predictions. What is the point of analysing the situation "*when compounds assigned to the 'both' class are considered correctly classified*"? If in practice we apply MCP and we get X % of "both" predictions in our test set, what will happen if they are considered correctly classified? First, as we are dealing with predictions we cannot know in advance their correct classification. Second, an obvious answer is that we would get X % more correct classifications! The improvements to specificity and sensitivity would depend on the proportion of positives and negatives in the test set, and that would then reflect the improvement of the CCR. Maybe there is a rationale or a practical benefit when analysing the situation "*when compounds assigned to the 'both' class are considered correctly classified*", but we do not see them nor have the authors provided an explanation for them.

### 'Empty' predictions

Our understanding is that the 'empty' predictions are treated as false predictions in the six metrics introduced by the authors. We totally agree with the authors' statement on page 7 that "*it is reasonable to argue that a compound assigned to the 'empty' class is too dissimilar from the molecules in the model and so cannot be part of the AD.*" Therefore, we are puzzled as to why they have not examined the option of removing 'empty' predictions from the denominator in their metrics. As 'empty' predictions are a fundamental part of CP we are also surprised to see that they are not even mentioned in the conclusion of the paper.

## Missed opportunities in the data analysis

Our main impression of the paper is a genuine regret for missed opportunities. Some of the issues discussed above we see as opportunities which, due to possible time or computer power constraints, might not have been feasible for the authors to execute. However, in our opinion there were options in the data analysis which would have brought additional value to the study with little extra effort. As mentioned before, the authors have treated 'both' predictions as either all being true (_incl metrics) or all false (_excl metrics), while 'empty' predictions have been treated as all being false. We regret not seeing the following options analysed in the paper.

### Removing uncertain predictions from the analysis

Near the end of the Temporal validation section of the paper (page 13) the authors have discussed grouping

'both' and 'empty' into a single category called uncertain and they illustrated this in Table 5. They even mention that "*Ignoring these predictions allows one to improve the overall predictivity*" (last paragraph page 13). There are indications (not generalisations) that by using a simplistic method of defining Uncertain predictions [3], different from the CP, the overall accuracy of binary predictions may be improved on samples which are not Uncertain. We regret not seeing that option analysed in the paper. To be more precise, we would have liked to see the results for following three statistics, and we believe that it would have improved the quality of the paper:

Specificity_uncertain_out = TN/(TN + FP)

Sensitivity_uncertain_out = TP/(TP + FN)

CCR_uncertain_out = (specificity_uncertain_out + sensitivity_uncertain_out)/2

Note: TN, TP, FN and FP values are as specified in our Table 1.

We would then be able to see if and to what extent statistics have improved at each confidence level (70%, 80%, 90%) and to compare that to the proportion of uncertain predictions (available from Table 2 in the paper). As such an analysis was not performed, we do not know the results. However, it would have been interesting, at least for us, to see if and to what extent the predictivity was improved on the subset of compounds predicted as positive and negative (not uncertain).

### Removing 'empty' predictions from the analysis with various scenarios for 'both'

In our view it might have been worthwhile investigating scenarios where 'empty' predictions are removed from the analysis and various options for 'both' predictions are examined. For example, 'both' classes may be considered as all active, or as all inactive, or as all equal to the dominant class in the training dataset. In short, we are of the opinion that there were more options for analysing 'empty' and 'both' predictions than those that the authors have presented us with.

### Comparison of the QSAR and MCP models

In the section Comparison of QSAR and Mondrian CP models of the paper (pages 10–11), as well as in the section Temporal validation (pages 12–13), approximately half of the MCP model performances are presented with the _incl statistics, i.e. "*when compounds assigned to the 'both' class are considered correctly classified*". As we have mentioned before, we do not understand the rationale for its use, because we cannot know in advance their correct classification. Furthermore, adding correct classifications cannot worsen the statistics, and the extent of improvement depends on the quality of the remaining predictions (_excl statistics). Unfortunately, the analyses of such situations lead to statements of the obvious. Here is just one example from the end of page 12.

"*The CCR is also affected whether or not the 'both' class predictions are considered when a confidence level of 90% is used. At this level, the CCR for the models including the 'both' prediction class reaches 0.86 compared with 0.43 when it is excluded. The greater number of compounds assigned to the 'both' prediction class at this confidence level results in globally better predictivity of the models (Table 4).*"

We are not stating that the authors are wrong here. The more correct classifications we have in our prediction sample the better its predictivity will be. However, is this not self evident?

### Conclusion

We disagree with the generalisation of findings as comparisons of QSAR and MCP models that is prevalent throughout the paper. Results presented are related to RF model with fixed parameters. We do not think that the design of the experiment in the paper is wrong as such, but the authors have not informed readers of its weaknesses. If the issues of variabilities could not have been addressed in the design of the study, due to possible time or computer power constraints, we would have liked to see them at least acknowledged in the paper. An important section of the paper related to model validation is not clear to us at all, therefore we are not able to repeat it. We again apologise in advance if the problematic section is clear to other people. We question the practical value of the way the authors have analysed 'both' and 'empty' predictions, and we provide cases for missed opportunities in their data analysis. We question the rationale behind half of comparisons between MCP and QSAR presented in the paper, because they examine the situations "*when compounds assigned to the 'both' class are considered correctly classified*". How can someone in practice transform 'both' predictions into correct classifications? How can it be useful in science to examine situations in which we assume that we know something which we cannot know?

As we see it, every research analysis (ours included) is bound to have weaknesses, and discussions regarding generalisation of results are common throughout science. We can imagine that Bosc et al. [1] have put a lot of effort in producing the results, and with all its inevitable weaknesses there was an opportunity, in our view, to provide some valuable indications (not generalisations) regarding the comparison of QSAR and MCP models. We do not think that the opportunity was used to its fullest potential. Taking everything into consideration, we think the approach taken by Bosc et al. [1] when comparing QSAR and MCP modelling methods is too narrow and methodologically unsound. Therefore, in our opinion, the scientific evidence presented by Bosc et al. [1] is not adequate to reach any conclusion whatsoever.

We would like to point out that Dr Nicolas Bosc, the corresponding author of the paper, was very helpful in replying to our emails and without his further clarifications of the _excl and _incl statistics our opinion piece would not have been the same.

### Acknowledgements
The author would like to thank his mother, Linda Louise Woodall Krstajic, for correcting English typos and language improvements in the text.

### Authors' contributions
The author read and approved the final manuscript.

### Funding
No funding received.

### Competing interests
The author declares no competing interests.

Received: 18 June 2019   Accepted: 14 October 2019
Published online: 06 November 2019

### References
1. Bosc N, Atkinson F, Felix E, Gaulton A, Hersey A, Leach AR (2019) Large scale comparison of QSAR and conformal prediction methods and their applications in drug discovery. J Cheminform 11(1):4
2. Krstajic D, Buturovic LJ, Leahy DE, Thomas S (2014) Cross-validation pitfalls when selecting and assessing regression and classification models. J Cheminform. 6(1):10
3. Krstajic D, Buturovic L, Thomas S, Leahy DE. Binary classification models with "Uncertain" predictions. arXiv preprint arXiv:1711.09677. 2017

**Publisher's Note**
Springer Nature remains neutral with regard to jurisdictional claims in published maps and institutional affiliations.